\documentclass[12pt]{article}
\input epsf.tex
\usepackage{cite}
\usepackage{latexsym}
\usepackage{graphicx}
\usepackage{amsmath,amstext}
\usepackage{comment}
\numberwithin{equation}{section}
\newcommand{\be}{\begin{equation}}
\newcommand{\ee}{\end{equation}}
\newcommand{\benn}{\begin{equation*}}
\newcommand{\eenn}{\end{equation*}}
\newcommand{\bea}{\begin{eqnarray}}
\newcommand{\eea}{\end{eqnarray}}
\newcommand{\bean}{\begin{eqnarray*}}
\newcommand{\eean}{\end{eqnarray*}}

\def\centeron#1#2{{\setbox0=\hbox{#1}\setbox1=\hbox{#2}\ifdim
\wd1>\wd0\kern.5\wd1\kern-.5\wd0\fi
\copy0\kern-.5\wd0\kern-.5\wd1\copy1\ifdim\wd0>\wd1
\kern.5\wd0\kern-.5\wd1\fi}}
\def\ltap{\;\centeron{\raise.35ex\hbox{$<$}}{\lower.65ex\hbox{$\sim$}}\;}
\def\gtap{\;\centeron{\raise.35ex\hbox{$>$}}{\lower.65ex\hbox{$\sim$}}\;}

\begin{document}
\begin{titlepage}
\begin{center}
\hfill SCIPP-2008/RUNHETC \\

\vskip 0.2in

{\Large \bf Holographic space-time from the Big Bang to the de
Sitter era}

\vskip 0.3in

Tom Banks ,$^1$

\vskip 0.2in

\emph{$^1$ Santa Cruz Institute for Particle Physics,\\
     Santa Cruz CA 95064 and Rutgers NHETC, 126 Frelinghuysen Rd., \\ Piscataway,
     NJ, 08854}

\begin{abstract}
 I review the holographic theory of space-time and its applications
 to cosmology. Much of this has appeared before, but the discussion
 is more unified and concise. I also include some material on work
 in progress, whose aim is to understand compactification in terms of finite
 dimensional super-algebras. This is an expanded version of a lecture I gave at
 the conference on Liouville Quantum Gravity and Statistical
 Systems, in memory of Alexei Zamolodchikov, at the Poncelet
 Institute in Moscow, 21-24 June, 2008.
.
\end{abstract}

\end{center}
\end{titlepage}

\section{Introduction}

\subsection{For Alyosha}
This paper was first prepared as a lecture to be delivered at the
conference honoring the memory of the great mathematical physicist
Alexei Zamolodchikov. Alyosha's untimely death was a shock and a
tragedy for many of us around the world, but especially for my good
friend, Alyosha's brother Sasha.  I was honored to be invited to
speak at this conference, and I will remember it for a long time.
Alyosha was a great physicist and a great man, and his friends gave
him the only kind of send off such a man deserves: a celebration of
his science and his life.

\subsection{Holographic space-time}
The holographic theory of space-time is an attempt to construct a
general theory of quantum gravity, which will include known string
theory models as special cases.  It is more flexible and local than
the existing formulation of string theory, and its general
principles are {\it background independent}\footnote{Actually this
phrase is somewhat misleading.  It implicitly views quantum gravity
as some sort of path integral over geometries, with different
backgrounds arising as stationary points of the integral.  As we
will see, this is completely wrong in the holographic theory.
Geometry arises instead as a collective variable of a system whose
fundamental formulation does not involve summing over geometries. In
particular, the quantum analogs of the causal structure and
conformal factor of a given holographic model are completely fixed
by its kinematics. The quantum variables are orientations of pixels
on causal diamonds.}. However, the holographic formalism immediately
reveals the different nature of the variables for space-times with
different asymptotics. The fact that the dynamical formulation of
string theory depends on the space-time asymptotics has been an
uncomfortable, but nonetheless valid conclusion that many string
theorists have drawn from existing models.

A second advantage of the holographic formalism (in this author's
eyes at least) is that it makes an immediate connection between
supersymmetry and the structure of space-time.  Indeed, the
fundamental quantum variables, which are interpreted geometrically
as the orientations of pixels on a holographic screen, can also be
viewed as the degrees of freedom of supersymmetric particles
penetrating the screen.  More precisely, they become supersymmetric
particle variables in the limit of large area screens. In this
formalism, it is impossible to make a Poincare invariant theory,
which does not contain the degenerate superpartners of all particles
in the theory.

The holographic formalism also sheds some light on the question of
{\it moduli stabilization}, which has haunted much of the history of
string theory.  In this formalism, a local description\footnote{Here
local refers to a description involving a single pixel of a finite
area causal diamond in the non-compact space.} of any compact
manifold, always assigns a finite dimension to its algebra of
continuous functions\footnote{In fact, there is no distinction
between measurable, continuous, and smooth functions at the local
level. The different function spaces arise as different limits of
the same sequence of finite dimensional function algebras.}. The
most attractive way to do this for even dimensional manifolds is to
assume they have a symplectic structure and construct the geometry
by geometric quantization, thus assigning them a finite dimensional
non-commutative algebra of functions.  We will see that this kind of
fuzzy compactification is also an appropriate way to think about the
geometry of the holographic screen of a finite causal diamond in the
non-compact dimensions. From this point of view then, moduli take on
a sequence of discrete values, and continuous moduli can result only
from infinite limits in space-times which admit causal diamonds of
arbitrarily large area. In particular, the hypothesis\cite{tbwfN},
which we will review below, that the quantum theory of de Sitter
space has a finite dimensional Hilbert space, already implies that
compact extra dimensions will have stabilized moduli, if space-time
is asymptotically de-Sitter in the future.

\section{Holographic space-time}

The basic building block of holographic space-time is not a point,
but a quantum causal diamond.  This is the quantum gravity
construct, to which a geometrical causal diamond is a classical
approximation.  A geometrical causal diamond is the intersection of
the interior of the forward light-cone of a point $P$, with that of
the backward light-cone of a point $Q$ in the causal future of $P$,
in some Lorentzian space-time.  The holographic screen of such a
diamond is the maximal area space-like $d-2$ surface on its
boundary.  The covariant entropy bound\cite{fsb} assigns a finite
entropy to a diamond with a finite area screen.  Fischler and
I\cite{holocosm} interpreted this as the entropy of the maximally
uncertain density matrix, the logarithm of the dimension of the
diamond's Hilbert space.

A pixel on the holographic screen, is an element of a basis in the
associative algebra of functions on the screen. We will assign a
finite dimensional Hilbert space to each pixel, and therefore a
finite area screen must correspond to an algebra with a finite
basis. We will allow it to be non-commutative, since this gives us a
much more efficient, systematic, and symmetric way to approximate a
classical space.  The full Hilbert space of the diamond is the
tensor product of the single pixel Hilbert spaces.

The single pixel Hilbert spaces are constructed using the idea that
a Lorentzian geometry can be encoded in the orientations in the
ambient space-time of all of the pixels (thought of in a naive
geometrical way), as well as the holographic areas, of a
sufficiently rich set of causal diamonds. At the classical level,
the orientation of a pixel is determined by a null ray and a screen
element transverse to it. Precisely this information is contained in
a solution of the Cartan-Penrose equation

$$0 = \bar{\psi} \gamma^{\mu} \psi (\gamma_{\mu})_{\alpha}^{\beta}
\psi_{\beta} ,$$ where $\psi_{\alpha}$ is a Dirac spinor. The vector
Dirac bilinear is a null-vector, and the solution of this equation
is a light-front spinor, which determines a transverse plane.  The
C-P equation is invariant under Lorentz transformations.  This local
Lorentz invariance is broken to transverse rotations by choosing a
gauge in which the light-front spinor $S_a^A$ occupies only the
upper components of the Dirac spinor.  The classical re-scaling
invariance of the C-P equation will be broken to a local $Z_2$ by
our quantization rule. Physically this means that quantum mechanics
introduces an area for each pixel, while the classical C-P equation
only determines an orientation.  The local $Z_2$ will be identified
with $(-1)^F$, which appears to be an exact gauge symmetry of all
known consistent string theory models. This choice automatically
builds the spin-statistics connection into all holographic
space-time models.   The local $Z_2$ also allows us to make a Klein
transformation, such that the mutually commuting variables
associated with independent pixels, become mutually anti-commuting.

With this preface, we can write down our ansatz for the commutation
relations

$$[S_a^A (m), S_b^B (n)]_+ = \delta_{ab} M^{AB} \delta_{mn} .$$ The
labels $m,n$ refer to a basis in the algebra of functions on the
holographic screen, so that our variables live in the spinor bundle
over the screen.   Small latin letters refer to spinors in the $d$
non-compact dimensions of space-time\footnote{It is worth pointing
out that in this formalism, de Sitter space is thought of as the
maximal causal diamond of a single observer, which is a non-compact
space, with boundary.}.  We are assuming that these non-compact
dimensions have at least an $SO(d-2)$ asymptotic symmetry, and in
fact we will assume a larger asymptotic symmetry group later.

The operators $M^{AB}$ live in the space of forms at a point on the
internal manifold. We will take the dimension of this manifold to be
$11-d$, anticipating a connection to supergravity.  It's natural to
suppose that the operators corresponding to forms of a given degree,
$p$ can be written in terms of sums over more primitive operators,
corresponding (in the geometric limit) to independent p-cycles on
the internal manifold, so that all of these operators have the
interpretation of wrapped p-brane charges. These independent p-cycle
operators, and the pixel operators $S_a^A$ will form a closed finite
dimensional super-algebra.  At the moment, all we require of the
super-algebra is that it have a finite dimensional unitary
representation.

In a compactification, the pixel label $n$ naturally has a tensor
factorization, corresponding to the tensor factorization of the
algebra of functions on space-time.  The algebra of functions on the
compact manifold will always be a finite matrix algebra, so we can
enlarge the operators $S_a^A$ and $M^{AB}$ by tensoring in these
matrices.  Once this is done, the internal geometry, to the extent
that it has meaning in the quantum theory, will be encoded in the
representation of the super-algebra generated by the brane charges
and the pixel operators.  String duality has taught us that internal
geometries are not absolute concepts in string theory.  As we change
their moduli (in cases where moduli exist), inequivalent topologies
can morph into each other, passing through regions where no
geometrical description exists.   It is only the conserved brane
charges which are defined everywhere on moduli space.  In a
holographic space-time, compact geometry is simply defined in terms
of the algebra of brane charges and pixel variables.  A classical
geometrical interpretation will be valid only in cases which have
moduli, and in extreme limits of the moduli space.

One striking feature of this formalism is that the number of
``functions" on the compact space will be finite for any finite area
causal diamond.  We will see the a convenient way to think of this,
for many of the spaces that arise in string theory compactifications
is to use {\it fuzzy geometry}.  The point is that many of these
spaces are compact Kahler manifolds, or can be thought of as one
dimensional bundles over a Kahler manifold.   Geometric quantization
allows us to view the algebras of functions on these spaces as
limits of sequences of finite matrix algebras.  We will discuss some
of the striking conceptual consequences of this observation in
section III.

\subsection{The Hilbert spaces of single observers and their
intersection}

In Lorentzian space time with no closed time-like curves, an
observer is modeled by a time-like world line whose tangent vector
is everywhere future directed.  In quantum mechanics the term
observer refers to a large system whose internal dynamics are well
approximated by a cut-off quantum field theory in a volume large
compared to the cut-off scale. The pointer variables of this
observer are averages of local fields over large volumes.
Measurements consist of dynamical entanglement of microscopic
variables with the large ensembles of states corresponding to fixed
values of the pointer variables.  Such measurements destroy quantum
coherence up to small corrections of order $e^{ - V}$ where $V$ is
the volume of the pointer in cutoff units.

In the real world, any such measuring device will have a mass and
travel on a time-like world line. In holographic space-time we model
such a world line as a nested sequence of causal diamonds whose tips
have larger and larger time-like separation along the world line.
For small enough time-like separation, the holographic screens of
these diamonds will have finite area. In space-times with an
asymptotic causal structure like that of Minkowski or de Sitter
space, the screen area will be finite for all finite time-like
separation\footnote{In dS space the area remains finite for any
time-like separation.}, while in space-times like AdS, the area goes
to infinity at finite time.  In all space-times with both
non-singular past and future\footnote{We will use the phrase {\it
scattering space-times} to describe space times whose asymptotic
past and future are both non-singular.}, it is convenient to view
the nested diamonds of a single observer to be centered about a
single point on the observer's trajectory. In Big Bang space-times
we instead take a sequence of diamonds whose past tips all lie on
the Big Bang singularity.  The latter rule will allow us to
incorporate the notion of {\it particle horizon} into our formalism.

 The quantum translation of these statements, whose validity we assert even
in regimes where no sensible geometrical picture exists, is a
sequence of Hilbert spaces ${\cal H} (t, {\bf x})$, such that ${\cal
H} (t, {\bf x}) = {\cal P} \otimes {\cal H} (t-1, {\bf x})$.  The
label ${\bf x}$ indicates a position on a $d_S = d - 1$ dimensional
spatial lattice.  This lattice determines the topology (but not the
geometry) of a space-like slice ${\bf S}$ {\it in the non-compact
dimensions}. As noted above, the topology of the compact dimensions
is not an invariant concept.  The compact invariants are the brane
charges, which appear in the pixel super-algebra.  They take on
topological meaning only when there are extreme regions of moduli
space. ${\cal P}$ is the representation space of the pixel
super-algebra. In a cosmological situation, the change of properties
of the compact space with cosmological time, would be encoded in $t$
dependence of the super-algebra.

The space-like slice ${\bf S}$ should be thought of as the
cosmological initial surface for Big Bang space-times.  For
scattering space times each observer is, loosely speaking, described
by a sequence of causal diamonds centered on some point. The
``initial" space-like slice is the one which goes through the
central points of all the observers on the lattice. For such
space-times, we have, instead of Hamiltonian dynamics, better and
better approximations to the scattering matrix, in terms of matrices
that compute outgoing from incoming data, in finite causal diamonds.

It should be noted that in situations where a geometrical
description is accurate, the integer variable $t$ is a monotonic
measure of the proper time $\tau$ traversed between the past and
future tip of the diamond represented by ${\cal H} (t, {\bf x})$,
but they are not linearly related.   In situations where the
internal geometry is unchanging, and the external geometry is weakly
curved, each increment in $t$ represents an equal increase in the
area of the diamond, so $t \propto \tau^{d-2}$. This implies that as
causal diamonds get bigger, the proper time is discretized in
smaller and smaller units.  Roughly speaking, the smallest proper
time interval measurable in a large causal diamond is inversely
proportional to the energy of a black hole whose horizon area is
equal to the area of the holographic screen.

The rest of the kinematical specification of a holographic quantum
geometry consists of a set of overlap rules.  We specify that
$${\cal H} (t,{\bf x}) = {\cal O} (t, {\bf x,y}) \otimes {\cal N} (t,
{\bf x,y}) .$$
$${\cal H} (t,{\bf y}) = {\cal O} (t, {\bf x,y}) \otimes {\cal N} (t,
{\bf y,x} ).$$  Note that the overlap Hilbert space ${\cal O}$ is
the same for ${\bf x}$ and ${\bf y}$, but the Hilbert spaces ${\cal
N}$ may be different. For nearest neighbor points on the lattice, we
insist that
$${\cal O} (t,{\bf y,x}) = {\cal O} (t,{\bf y,x}) = {\cal P},$$ for
all $t$. Geometrically, this is the requirement that the
trajectories of nearest neighbor observers always share all but one
pixel's worth of information. The dimension of the overlap Hilbert
space is required to be a monotonically non-increasing function of
the minimal number of lattice steps between  ${\bf x}$ and ${\bf
y}$.  The rest of its specification is part of the dynamics of the
system.

To discuss the dynamics, we must make some specification of the
space-time asymptotics.   We distinguish three cases

\begin{itemize}

\item Space-times, like dS space, in which there is a maximal area causal diamond.

\item Space-times, like Minkowski space, where the area of causal
diamonds goes to infinity continuously as a function of the proper
time in the diamond.

\item Space-times, like anti-deSitter space, in which the area
variable $t$ goes to infinity at a finite value of the proper time.
Conformal infinity in such space-times is timelike.

\end{itemize}

We can find examples in the first two categories of both Big Bang,
and scattering space-times, while in the third category I only know
of scattering space-times. The AdS/CFT correspondence suggests that
the proper formulation of the quantum dynamics of the third category
is in terms of a quantum field theory living on the conformal
boundary of space-time. To be more precise, we must insist that the
conformal boundary be identical to that of Anti-de Sitter space. The
rate of approach to the background metric determines whether the QFT
is conformally invariant, or is a relevant perturbation of a
conformal field theory\footnote{It should be noted that the
conformal boundary is taken to be a spatial sphere cross time and
the Hamiltonian operator in the far UV of the field theory is the
standard conformal generator $K^0 + P^0$.}.  The compact dimensions
of the bulk space-time are encoded in the target space of the field
theory, rather than in the space-time geometry on which the field
theory lives.  This is analogous to the separation of compact and
non-compact dimensions in our general theory of holographic
space-time.

Local physics in causal diamonds much smaller than the AdS radius is
not easy to disentangle from the field theoretic dynamics on the
boundary.  There are several proposals to deal with this
issue\cite{adslocality}, none of them fully satisfactory.

One can also imagine deformations of these field theories in which
the exactly marginal and relevant couplings of the boundary theory
are allowed to violate the symmetries of the field theory, including
time translation invariance. In principle one could try to model
cosmology on field theories with time dependent couplings, but the
time dependence seems quite arbitrary. Furthermore one would have to
specify the initial state of all degrees of freedom in the system,
and since the Hamiltonian couples them all at all times, it is not
clearly why such a ``cosmology" would have particle horizons or
sensible local physics.

Instead, Fischler and I proposed\cite{holocosm} to attack the
problem of cosmology directly within the holographic formalism. Our
mathematically well defined holographic cosmology is called {\it the
dense black hole fluid} (DBHF). It is defined by the following set
of rules:

\begin{itemize}

\item Each observer on the spatial lattice is given the same
sequence of time-dependent Hamiltonians $H(t)$\footnote{The dynamics
is discrete and $H(t)$ is just $i$ times the logarithm of the
unitary transformation which transforms the system from $t$ to
$t-1$. The parameter $t$ goes from $1$ to $\infty$.  All of the
operators $H(t)$ operate in the late time Hilbert space ${\cal
H}(\infty , {\bf x})$.}

\item The operator $H(t)$ is the sum of an operator $H_{in} (t)$
which is built from the pixel operators operating in ${\cal P}^t$
and an operator $H_{out} (t)$ which commutes with all of those
operators.  This rule builds the concept of particle horizon into
the dynamics.

\item $H_{in} (t)$ is a perturbation of a bilinear Hamiltonian in
the fermionic pixel variables.  The bilinear form is chosen
independently for each $t$ from the Orthogonal ensemble of
anti-symmetric $t \times t$ matrices, with a simple $t$ dependent
normalization. For large $t$ this bilinear Hamiltonian describes a
scale invariant free fermion in $1+1$ dimensions.  The perturbations
are required to be irrelevant perturbations of this CFT.

\item  The overlap rule is ${\cal O} (t, {\bf x,y}) = \bigotimes {\cal P}^{t -
d({\bf x,y})}$, where $d({\bf x,y})$ is the minimal number of
lattice steps between the two lattice points.  The Hamiltonians on
the overlaps are just $H(t - d({\bf x,y}))$.  As a consequence of
the fact that we required the sequence of Hamiltonians to be the
same at each lattice point, this rule satisfies all of the
complicated consistency conditions.

\end{itemize}

One can then show that the system has an emergent geometry, which is
a spatially flat FRW universe with $p=\rho$.  To give some feeling
for how this works, the overlap rule gives a formula for the {\it
causal horizon}, the boundary between the set of lattice points with
which a given observer has interaction, which is defined by a rule
that becomes rotationally invariant at large $t$, for any regular
lattice with the topology of ${\bf R}^3$.  The size of this causal
horizon scales with $t$, interpreted as the area of the causal
diamond reaching back to the big bang, as expected for the FRW
geometry.  The energy density is defined in terms of the Hamiltonian
of the system, and also obeys the right scaling law, as well as the
relation $\sigma = k \rho^{1/2}$, between entropy and energy
densities, expected for a $p=\rho$ fluid with extensive entropy.
This is also the relation between entropy and energy density for a
``system of horizon filling black holes", which constantly merge to
fill the growing horizon.  The phrase in quotes is just an heuristic
description of the actual mathematics, but it gives rise to the name
DBHF that we have given to this system.

The utility of this phrase comes from considering a cosmology
consisting of a dilute gas of black holes, with a sufficiently
homogeneous distribution. It is clear on the one hand that locally
the black holes are regions of space-time packed with the maximum
allowed entropy, but that on the other hand, this system behaves
like an FRW universe with $p = 0$ for a very long time.  However,
fluctuations in such a universe grow with scale size and we
eventually have to ask what will become of the system.  Clearly,
fluctuations will lead to larger and larger black holes through
collapse and collision. Eventually, perhaps depending on the initial
density and fluctuations, we might expect the system to behave like
the DBHF.

Quite remarkably, the opposite phase transition from dense to dilute
black hole fluid can also occur. This observation is the basis of
the realistic cosmology that Fischler and I proposed\cite{holocosm}.
It begins with the heuristic idea of a {\it defect} in the DBHF.
Geometrically this is a region of coordinate space, in the flat FRW
coordinates defined by the DBHF, in which the dynamics of the system
leads to a lower entropy initial configuration.  We may imagine that
in certain regions we have initial black holes, which are too small
to merge with their neighbors in adjacent horizon volumes, as the
universe expands.  These would decay into particles\footnote{The
right way to think about particles in the holographic space-time
formalism will be adumbrated below.} and this {\it normal region of
space-time} will expand locally like a radiation dominated FRW
universe.

For large $t$, the system also has a scale invariance, that of the
massless $1+1$ dimensional fermion field. This can be shown to be
the same transformation that implements the conformal isometry $\tau
\rightarrow c \tau$, ${\bf x} \rightarrow c^{ {2\over 3}} {\bf x}$
of the emergent geometry
$$ds^2 = - d\tau^2 + \tau^{2/3} (d{\bf x} )^2 .$$

The holographic formalism defines a natural time slicing for Big
Bang universes, whether spatially homogeneous or not.  We simply
insist that the integer $t$, which defines the time slice, refers to
the area of the causal diamond associated with the local particle
horizon: the intersection of the interior of the backward light-cone
of the observer at ${\bf x}$ with the interior of the forward
light-cone of the Big Bang event at ${\bf x}$.  In the quantum
theory this is built in to the rule that ${\cal H} (t, {\bf x}) =
\otimes {\cal P}^t $. It is easy to see that, in such a slicing, a
normal region that {\it expands freely} has spatial volume growing
more rapidly than that of the DBHF.  Thus, the spatial volume
fraction of normal region relative to DBHF grows with time. As a
consequence, at some physical size $M$ for the cosmological horizon,
the normal patch of the universe looks like a radiation filled
universe interspersed with patches of DBHF.  The latter behave like
black holes: local regions of maximum entropy, with all of their
degrees of freedom in equilibrium.  Thus, we have achieved a
transition between the DBHF and a dilute black hole gas. If the
dilute black hole gas is approximately homogeneous this system will
quickly begin to evolve like a $p=0$ FRW universe, but well known
gravitational instabilities will bring it back to the DBHF very
quickly, unless the inhomogeneities are very small.  {\it Thus,
holographic cosmology can explain the low entropy of the initial
conditions for the normal part of the universe. One would hope to
eventually prove that any higher entropy of normal degrees of
freedom would lead to re-collapse to the DBHF}. A period of
inflation can help to avoid this instability, if the theory contains
a low energy effective inflaton field.  The holographic framework
provides the explanation for why initial configurations of this
field {\it must} be approximately homogeneous: again to avoid
re-collapse into the DBHF fluid. The reader should note carefully
that homogeneity, isotropy and flatness are, in this formalism, all
derived for generic initial conditions, without inflation. Indeed, I
do not believe that the convention {\it inflationary} explanation of
these conditions really explains anything, because it makes drastic
assumptions about the initial state of those degrees of freedom that
can not be approximately described by local field theory in the
initial inflationary patch. The real test of the explanation of low
entropy initial conditions from the principle of avoiding
re-collapse to the DBHF, would be a first principles calculation of
the amplitude of primordial density fluctuations. This does not yet
exist.

The picture sketched in the previous paragraph depended on the
dynamical assumption that the normal region could expand as if the
surrounding DBHF was not there. A rigorous investigation of this
requires a fully quantum and holographic description of the normal
region, and its interactions with the DBHF. However, we can obtain
interesting insights by looking at the implications of the Israel
junction condition for a spherical region of $p = w \rho$ cosmology
embedded into a $p = \rho$ background. We find that, applying the
condition that the geometry of the junction be continuous, the
coordinate volume of the $p = w \rho$ region shrinks unless $w = \pm
1$. In the de Sitter case we match the cosmological horizon, which
is a marginally trapped surface, to the horizon of a black hole
geometry of equal area in the $p=\rho$ background. This implies that
there is a stable equilibrium between an asymptotically de Sitter
normal region and the DBHF, which does not exist for any other
asymptotic FRW geometry.

I view this as a prediction of a positive cosmological constant,
whose value is determined by initial conditions. One can start with
many finite defects in the DBHF, each one involving some finite
number of the pixel variables\footnote{Recall however that we do not
yet have a description of the defects in terms of pixel variables.},
which would evolve to a {\it lonely multiverse} of isolated
asymptotically dS universes. I call it lonely, because there is no
reason for these universes to have any effect on each other.  It
would be an amusing problem in GR to determine the properties of a
solution with two trapped surfaces in a space-time asymptotic to the
flat $p = \rho$ universe. Do they attract, repel, collide? Whatever
the answer however, one can surely choose the initial distribution
of universes in such a way that the collision takes place only long
after each universe has reached its asymptotic dS regime. Thus,
depending on unknown initial conditions we could always arrange that
there is no observable effect of collisions even if they occur. We
might as well insist that no such collisions occur.

Returning to the Israel argument for stability, we note that the
entropy associated to the dS space coincides with that excised from
the $p=\rho$ universe, so the thermodynamic argument for stability
mirrors the geometrical one. Another way of putting this is that the
de Sitter vacuum is a state which maximizes the entropy in a given
causal diamond. In the DBHF, the degrees of freedom of this diamond
would mix with the other degrees of freedom in the universe, but by
modifying the metric around the de Sitter bubble to be that of the
``$p=\rho$ Schwarzschild solution" we find a stable solution of
general relativity, and thus, local thermodynamic stability. If we
treat fluctuations around this solution by the methods of quantum
field theory, we would find a Hawking instability, but our explicit
model of the DBHF makes it clear that it has no particle
excitations. Indeed, the coarse grained Friedman equations are
derived from a quantum model whose time dependent Hamiltonian is
constantly moving the state vector around in Hilbert space, rather
than a system with a unique time independent ground state. Our claim
is that the only stable ``excitations" of the DBHF are stable black
holes with de Sitter interiors.

 The Israel condition argument also implies that the initial
normal region must have a complex shape. If it were spherical, it
would be invaded by the DBHF before it reached its asymptotic dS
limit. To be more precise, we have two choices:
\begin{itemize}

\item We can assume the initial normal region is spherical, but takes up a
much larger coordinate volume (more points on the lattice) than our
current horizon volume does, so that even though the coordinate
volume shrinks in response to the pressure of the external $p=\rho$
region, our full horizon volume survives until de Sitter expansion
begins, {\bf or}

\item We assume a non-spherical shape, determined by maximizing the
initial fraction of the coordinate volume, which is in the DBHF
fluid phase, {\it subject to the constraint that the phase
transition to a more normal universe occurs, and that the normal
phase remains stable until de Sitter expansion takes over}. We have
argued that the de Sitter phase of expansion of the normal universe
is stable against re-collapse into the DBHF.

\end{itemize}
There is more entropy in the second kind of initial condition, and
indeed the first kind is simply a low entropy example of the second.
It is attractive to assume that the initial state of the universe is
as generic as possible, since we then feel no compulsion to explain
its properties. We have seen that the most generic initial
conditions lead to the DBHF, a universe in which nothing ever really
happens. Within the bounds of our ignorance about the detailed
mathematical formulation about the DBHF-defect model of the
universe, it seems reasonable to characterize it as the most generic
initial state that can lead to complex evolutionary behavior in the
future.

To summarize, holographic cosmology predicts that the normal region
of the universe must asymptotically approach de Sitter space, with a
cosmological constant determined by the cosmological initial
conditions.   $\Lambda$ is determined by the number of degrees of
freedom that were initially out of equilibrium with the background
DBHF. In other words, holographic cosmology is automatically a {\it
multiverse} theory.  Separated small defects in the DBHF evolve into
normal universes with a variety of values of the cosmological
constant.  Our own universe can thus be subject to environmental
selection effects. There is however an {\it a priori} preference for
the largest value of the c.c. that can be compatible with the
anthropic constraints, since that represents the maximum entropy
initial condition for the universe. The answer to the question of
whether other parameters in the Lagrangian describing low energy
physics are determined environmentally, depends on the degree of
uniqueness of the quantum theory of de Sitter space, a subject to
which we now turn.

\section{The quantum theory of stable de Sitter space}

\subsection{The two Hamiltonians of Wm. de Sitter}

We begin by recalling some semi-classical properties of de Sitter
space.  We will work in four dimensions, inside a single horizon
volume, which may be covered by a static metric

$$ds^2 = - f(r) dt^2 + {{dr^2}\over f(r)} + r^2 d\Omega^2 .$$
For the dS ``vacuum", $f(r) = (1 - (r/R)^2)$, while for the
Schwarzschild-dS black hole, $f(r) = (1 - {{2M}\over {r M_P^2}} -
(r/R)^2 )$.  Gibbons and Hawking \cite{GH} argued that both of these
metrics represented thermodynamic ensembles of states.  In
particular, the vacuum ensemble is canonical, with {\it unique}
temperature $T^{-1} = 2\pi R$, and entropy $S = \pi (RM_P)^2$.  The
black hole metric has two horizons, each with its own entropy and
temperature.  For $ M \ll M_P^2 R$, the inner horizon is
approximately that of a Minkowski black hole, while the outer
horizon is approximately that of the vacuum ensemble. It is
important that the combined entropy is {\it always} less than that
of empty dS space. For small $M$ the entropy deficit is $2\pi R M$.

The hypothesis that the entropy of dS space is just the logarithm of
the dimension of the Hilbert space in its quantum theory\cite{tbwfN}
leads to a natural interpretation of these facts. Remarkably, the
explanation invokes {\it two} different Hamiltonians. The first,
which we denote by $H$, is an operator with spectrum bounded by $c
T$, with $c$ a constant that has not yet been determined. It is
likely that it will turn out to be a ``random" operator, in that all
properties of dS space which are {\it in principle} amenable to
measurement will depend only on certain gross properties of $H$. The
first is that $H$ has a chaotic spectrum, so that a generic initial
state cycles through the entire Hilbert space under $H$ evolution.
This statement is meaningful only for a Hilbert space of very large
dimension, but that will certainly be the case for the Hilbert space
representing our own universe. We will see that the quantum theory
of dS space really only makes sense when its entropy is large. The
vacuum ensemble of Gibbons and Hawking will be identified with the
ensemble of all states of a Hilbert space of dimension $e^{\pi
(RM_P)^2 }$, evolving under the Hamiltonian $H$.

Given this identification, we can immediately understand another
aspect of semi-classical de Sitter physics.  The
Coleman-DeLucia\cite{cdl} instanton for transitions between two dS
vacua is a compact Euclidean manifold with negative Euclidean
action. It defines {\it two} transition probabilities upon
subtraction of the Euclidean actions of the two dS spaces.  These
describe inverse processes and the probabilities are related by
$$P_{12} = P_{21} e^{S_1 - S_2} .$$ This is the principle of
detailed balance for a system in equilibrium at infinite
temperature, which is precisely the interpretation of the dS vacuum
state we gave in the previous paragraph.

The Hamiltonian $H$ is certainly not the Hamiltonian whose
eigenvalues are particle masses in the real world, and the vacuum
ensemble is an infinite temperature ensemble for $H$. This indicates
the need for another Hamiltonian, $P_0$, to describe local physics
that is approximately Poincare invariant in the large $RM_P$ limit.
Indeed, the manner in which the cosmological causal diamond in dS
space approaches Minkowski space, indicates the existence of two
Hamiltonians.  Near the cosmological horizon, the static metric
approaches
$$ ds^2 = R^2 (- du dv + d\Omega^2),$$ where the horizon is the surface $v
= 0$. The metric of asymptotically flat space near conformal
infinity is
$$ds^2 = {1\over v^2} (- du dv + d\Omega^2).$$ The relation between
the two in the $R\rightarrow\infty$ limit is clear. The Lorentz
group is realized as the conformal group of the two sphere, and the
translation generators are
$$ P_{\mu} \propto (1, {\bf \Omega}) \partial_u .$$  By contrast,
the static Hamiltonian acts on the dS metric as
$$H \propto (u\partial_u - v\partial_v ),$$ so that
$$[H, P_{\mu} ] \propto P_{\mu} .$$

Global dS space does not have an infinite null or time-like
boundary, so it is not clear from the canonical formalism of gravity
how one should interpret its isometries.  One can argue that they
are all gauge generators, which should be set to zero.  However,
such a formal argument would apply to any finite causal diamond in
the holographic formalism. Instead, one should look at this
formalism is working in a fixed physical gauge, defined by some set
of physical measuring devices.

My current understanding of the quantum theory of measurement relies
on (cut-off) quantum field theory.  A measurement consists of a
dynamical entanglement of some microstate with the ensemble of
states corresponding to a fixed position of a pointer variable.
Pointer variables are averages of local fields over volumes large
compared to the cutoff.  Tunneling between states corresponding to
different pointer positions is suppressed by $e^{- V_{pointer}}$,
with the volume measured in cutoff units. Thus, I believe that the
only kind of states for which we have a reliable measurement theory
are those which are localizable.  They are either described by bulk
quantum field theory, or consist of black holes small enough to have
their states measured by devices that obey quantum field theory.

In dS space, all such states are evanescent and decay eventually to
the dS vacuum.  Our proposed model for these facts consists of the
dS Hamiltonian $H$ of a few paragraphs back, and an operator $P_0$
satisfying

$$[H,P_0] = M_P^2  g( {{P_0}\over {R M_P^2}}) ,$$ ({\it cf. the
commutator of generators on the cosmological horizon}) where $g(x)$
is a smooth function which is $o(x)$ for small $x$. The spectrum of
$P_0$ runs from $0$ to the (Nariai) mass of the maximal black hole
in dS space. For eigenvalues of $P_0$ small on the Nariai scale, the
entropy deficit of the eigenspace, relative to the full dS Hilbert
space, must be $2 \pi R P_0$.  This implies that the infinite
temperature ensemble for $H$ is the dS temperature ensemble for
$P_0$ as required by the match to semi-classical physics. Note that
this relation between entropy deficit and eigenvalue is {\it
precisely} what we observed above for black holes, if we identify
the black hole mass with an approximate eigenvalue of $P_0$.  We
will provide a somewhat more refined, but still crude, model of
$P_0$ in the third subsection of this chapter.

Returning to what we said about CDL transition probabilities, the
quantum interpretation of the CDL transition refers to the the
Hamiltonian $H$, rather than the emergent Hamiltonian $P_0$, whose
spectrum consists of states with lifetimes short compared to the CDL
transition time scale. While we are on the subject of CDL
instantons, it is worthwhile pointing out the distinction between
stable and unstable dS spaces, which appears in the CDL formalism.
If the potential contains a zero c.c. minimum, or an asymptotic
region where the energy density is zero, then dS space is unstable
to decay to the zero c.c. region. There is no inverse transition,
and this is interpreted by saying that the zero c.c. configuration
represents a system with an infinite number of states. If there is
no zero energy point on the potential, even at infinity in field
space, then the space of potentials divides into two
classes\cite{abj}. The distinction is based on the behavior of
tunneling amplitudes in the limit that the lowest dS minimum is
shifted to zero.  If the resulting Minkowski space has a positive
energy theorem, so that it is stable, then all ``decays" of the
lowest dS minimum, including those to negative c.c. Big Crunches
behave like $e^{ - \pi (RM_P)^2}$ for large dS radius, and can be
interpreted as improbable transitions to low entropy states of a
finite system (the lowest dS vacuum).  If the decay proceeds even in
the Minkowski limit, then the dS space is unstable, and it is not
clear whether the system has a sensible quantum mechanical
interpretation.

My claim that the negative c.c. crunches, must, in some cases,
transition back to empty de Sitter space, has caused some raised
eyebrows. What I want to emphasize is that this {\it must} be the
case if one assumes that the quantum system being described has a
number of states bounded by the exponential of one quarter the area
of the maximal sized causal diamond in the space-time. The reverse
transitions then follow from unitarity. Reference \cite{tbwfsing}
provides evidence that the singular CDL crunches indeed correspond
to sub-systems with microscopically small entropy.

\subsection{The variables of dS quantum mechanics}

In accord with our general formalism, the variables of a quantum dS
space satisfy the anti-commutation relations
$$[(\psi^a )_i^A , (\psi^{\dagger\ b} )_B^j ]_+ = \delta_i^j
\delta^A_B M^{ab} .$$  $a,b$ are 8 component spinor indices and the
superalgebra of $M$ and $\psi$ specifies the geometry of
compactified dimensions. The indices $i$ and $A$ run from $1$ to $N$
and $1$ to $N+1$ respectively and the $SU(2)$ rotation symmetry of
the cosmological horizon acts on these like a section of the spinor
bundle over the fuzzy 2-sphere.  The entropy of this system is
$N(N+1) {\rm ln} D$, where $D$ is the dimension of the
representation of the compactification superalgebra.  For large $N$,
this is identified with $\pi (RM_P)^2$, with $M_P$ the four
dimensional Planck scale. Thus $N \sim R M_P$.

I have not yet worked out the theory of fuzzy compactifications, so
I will follow \cite{bfm} and drop the internal spinor indices and
the brane charges $M^{ab}$. This leads to a Hilbert space containing
only chiral multiplets of minimal four dimensional SUSY, and no
graviton. Nonetheless, we'll be able to see how particles,
super-symmetry and black holes arise in the large $N$ limit.

Note by the way that we only try to construct dS space of dimension
four, where the holographic screen is a two sphere. This is
motivated by supergravity. For large dS radius, the quantum theory
should contain states which are well described by supergravity, and
for self consistency, the supergravity Lagrangian should have dS
solutions. The only examples I know,have at most four supercharges,
which implies four or fewer dimensions. The interesting physics in
dS space is the almost Poincare invariant physics of localizable
states. In fewer than four dimensions there is no sensible notion of
an S-matrix in the presence of supergravity.

The clue to the proper description of particle states in dS space
comes from a simple exercise first done in \cite{nightmare}. We ask
how to maximize the entropy described by quantum field theory in a
single horizon volume. In field theory, one maximizes entropy by
going to high energy. In a region of size $R$ the entropy of a field
theory will be of order $ (M R)^3 $, where $M$ is the ultra-violet
cut-off. The mass in this region is of order $M (MR)^3 $. The
Schwarzschild radius $M^4 R^3 / M_P^2 $ of this mass must be less
than $R$. Otherwise we are talking about black hole states, for
which the field theory description is incorrect. This leads to a
bound $M < (M_P / R)^{-1/2}$. Of course, this cannot be an absolute
bound on the momentum of any particle. Rather, it is a bound on the
momentum of the particle states of maximum entropy, which can exist
in dS space. Fewer particles of higher momentum would also evade the
black hole bound. Field theory does not provide a concise
description of how to describe such a restriction on the allowed
particle states. We will see that the holographic description
accomplishes this in an elegant manner.

The forgoing argument shows us how the formalism of quantum field
theory in curved space-time, in which there appear to be an infinite
number of copies of the degrees of freedom in a single horizon
volume at large values of the global time, might emerge as a limit
of the quantum dS formalism. The field theoretic entropy in a
horizon volume is of order $(R M_P)^{3/2}$. Consequently, the
Gibbons-Hawking entropy allows us to have of order $(RM_P)^{1/2}$
independent copies of these degrees of freedom. The field theory
prediction is recovered in the limit $(RM_P) \rightarrow\infty$.

The holographic description of these particle states is obtained via
the block decomposition of the fermionic matrix:

$$\psi_i^A = \begin{pmatrix} 1 & 2 & 3 & \ldots & K \cr K & 1 & 2 & \ldots
& K - 1 \cr \vdots & \vdots & \vdots & \ldots & \vdots \cr 2 & 3 & 4
\ldots 1 \end{pmatrix}$$ Each diagonal labeled by the same integer
represents the degrees of freedom in a single horizon volume, while
each block within that diagonal represents the degrees of freedom of
a single particle in that horizon volume. In order to have all
horizons equivalent, we must have $K \sim N^{1/2}$. If, following
the suggestion of Matrix Theory\cite{bfss} we now identify the
radial component of  momentum of a particle, in units of the dS
temperature with the trace of the block size, then our typical
momentum is of order $N^{- {1\over 2}}$ in Planck units, which is
the same as the momentum cutoff in our heuristic field theory
argument. Larger values of momentum can be obtained by lowering the
entropy.

To understand this we note that in the large $N$ limit, the
operators in each block of the single horizon volume diagonal,
converge to
$$ \psi \sqrt{p} \delta (\Omega - \Omega_0 ) ,$$ where $\psi
$ is a single fermionic annihilation operator, and $\Omega_0$ runs
over the two sphere. The block sizes all go to infinity with fixed
ratio and physics becomes invariant under rescaling of the block
sizes: this is Lorentz boost invariance under boosts in the
$\Omega_0$ direction\footnote{I should emphasize that throughout
this article, we are dealing with kinematics. The requirement that
the dynamical S-matrix be invariant under Lorentz transformations is
a strong constraint on the dynamics.}. As promised, the variables
are sections of the spinor bundle over the two sphere\footnote{To be
more precise: they are elements of the dual space to the space of
measurable spinor sections.}. The positive real number $p$ is the
re-scaled trace of the unit matrix in a single block.  It is
interpreted as the overall scale $p (1, \Omega_0 )$ of the lightlike
momentum of a massless superparticle, which exits the holographic
screen in direction $\Omega_0$. The rotation generators acting on
the rows and columns of the single $K\times K+1 $ block identify the
representation of SUSY as the chiral supermultiplet.  The SUSY
generators are

$$Q_{\alpha} = \psi q_{\alpha} (\Omega_0 ),$$ and their complex
conjugates.  $q_{\alpha} (\Omega_0 )$ is one of the conformal
Killing spinor of the two sphere, which transform like a left handed
Weyl spinor under the $SO(1,3)$ conformal group. The latter is, as
usual, identified with the Lorentz group, and we are working in a
basis where the single particle momentum is diagonal. Note that the
limit from dS space picks out a special Lorentz frame, the one in
which a particular static dS observer is at rest.

For finite $N$, the trace of a block is interpreted as the overall
scale of momentum, in units of $1/R$. Thus, for the maximally
entropic single horizon particle configurations where $K \sim
\sqrt{N}$, the momentum is of order $R^{- {1\over 2}} M_P^{1/2} \sim
\Lambda^{1/4} $.  We can obtain higher momentum by recognizing that
particles are here defined as in experimental physics: by their
imprints on the detector (the holographic screen).  Thus, $B$ blocks
which exit the same pixel (have the same angular momentum wave
function) will be interpreted as a single particle with momentum $B
\Lambda^{1/4}$.  Since $B$ can be as large as $N \sim (R M_P)$, the
momentum can be as large as $10^{30} M_P$. The physical origin of
this bound is a bit obscure. Any particle in $dS$ space will scatter
off the Gibbons-Hawking radiation, which would create a black hole
for high enough particle momentum. However, a typical GH quantum has
energy of order the dS temperature, so the center of mass energy in
the collision would be of order $\sqrt{ET}$ and it would seem that
the threshold for black hole production is $E \sim M_P (R M_P)$,
rather than $E \sim (RM_P )^{{1\over 2}} M_P$. We will see in the
next section that black hole states in a given horizon volume indeed
borrow particle degrees of freedom from other horizons and that
black holes are like particles with very high momentum, at least
insofar as their count of degrees of freedom is concerned. They
differ from particles in that they generically do not have
multi-black hole states related by permutation symmetry.

For the time being I will leave this small puzzle about the maximum
momentum unresolved, and proceed to the discussion of SUSY breaking.
We have seen that in the large $N$ limit, the full super-Poincare
algebra emerges. For generic particle states, with momentum of order
${{\sqrt{N}}\over R}$, the corrections to the Super-Poincare algebra
would be of order $N^{ - {1\over 2}} \sim (R M_P)^{- {1\over 2}}
\sim ({\Lambda \over {M_P^4}})^{1/4}$. The corrections to the
commutator of $Q_{\alpha}$ and $P_0$ should be of this order,
measured in Planck units. Recalling that in the super-Higgs
mechanism, the superpartner of any state is a state with an
additional massive gravitino, we get the prediction
$$m_{3/2} = \kappa \Lambda^{1/4} .$$ Interestingly, this coincides
with an heuristic estimate\cite{susyhor}, which we review in the
appendix, based on a model of gravitino interactions with a random
system spread over the horizon with a uniform density of states.

The estimate of the mass scale of standard model superpartners,
which follows from combining this equation with gravitational
effective field theory is $M_S = \sqrt{{\kappa \Lambda^{1/4} M_P}
\over {(8\pi)^{1/2}}} = \sqrt{2\sqrt{10}\kappa} TeV$. This is of
course as low as it could possibly be and still be consistent with
current experimental bounds. The precise estimate depends on the
unknown constant $\kappa$.

The strategy I have adopted for pursuing the phenomenological
consequences of these ideas is based on the fact that holographic
cosmology implies that $N$ is a cosmological initial condition, and
is therefore a freely variable input parameter. In the large $N$
limit, SUSY breaking is a very low energy phenomenon and, apart from
the question of what fixes the cosmological constant, it should be
understood in low energy effective field theory. In particular, the
$N=\infty$ theory should be supersymmetric and R symmetric (to
explain the vanishing of $\Lambda$ in the limit). The R symmetry
will be discrete, in accord with general ideas about global
continuous symmetries in gravity. For finite $N$, the Lagrangian
will contain $R$ breaking terms, the nature of which can only be
computed from the full quantum theory of dS space. These must induce
a SUSY violating vacuum state. The {\it size} of the R-breaking
interactions is determined by the requirement that the gravitino
mass, a quantity which can be calculated reliably in local field
theory is of order $\Lambda^{1/4}$. A constant term in the
superpotential is added to make sure that the c.c. is of order
$\Lambda$. We do not worry about the fine tuning required for this
parameter, because it is implementing a property that we know to be
imposed by the full quantum theory. The most phenomenologically
successful implementation of this strategy is called The Pentagon
Model, and is reviewed in \cite{pentagon}.

Returning to the underlying quantum theory, we note that different
blocks of the same size are related by gauge transformations: change
of basis in the $\psi_i^A$ index space. It is important to note that
whereas the $i$ and $A$ indices are chosen to transform under the
rotations of the cosmological horizon, this is not the case for the
emergent Lorentz group in the $N \rightarrow\infty $ limit. Instead,
rotations act on the indices within individual blocks, whereas the
transformations that exchange blocks are viewed as gauge
equivalences. We have seen above that the Hamiltonian $H$ is
distinct from the emergent Hamiltonian $P_0$, which acts on
localizable states in a single horizon volume, and becomes a
generator of the Poincare group in the large $N$ limit. Now we see
that the generators of $SO(3)$ rotations in dS space are not the
same as their Poincare limits.

It's amusing to note that the unitary transformations which exchange
diagonals are also viewed as gauge transformations in this
formalism. These are discrete analogs of elements of the $SO(1,4)$
dS group. It has been suggested\cite{arvindlowe} that the quantum
theory of dS space might be invariant under a q-deformed version of
the dS group, in order to be consistent with a finite dimensional
Hilbert space. Perhaps these discrete gauge transformations should
be thought of in this light.

At any rate, it has become abundantly clear that the local physics
in dS space has little to do with generators which act globally on
the space.  Instead it is encoded in the emergent super-Poincare
group of the $R\rightarrow\infty$ limit.

\section{Black holes in dS space}

The metric of a Schwarzschild-de Sitter black hole is

$$ds^2 = - f(r) dt^2 + {{dr^2}\over {f(r)}} + r^2 d\Omega^2 ,$$
where $$ f(r) = 1 - R_s/r - (r/R)^2 .$$  The horizons are at the
positive real zeroes of $f(r)$ and satisfy

$$R_+ R_- (R_+ + R_-) = R_s R^2,$$ and $$R_+ R_-  + R_+^2 + R_-^2 =
R^2 ,$$  where $R_s = 2M/M_P^2$, is the Schwarzschild radius of a
black hole of mass $M$ in asymptotically flat space-time. In accord
with the notation, we insist that $R_+
> R_-$. The second of these equations shows that the combined
entropy of black hole and cosmological horizons is less than that of
empty dS space. This leads us to the conclusion that localized
excitations all have an entropy deficit: they are not typical states
in the dS vacuum ensemble (which we recall is the infinite
temperature ensemble for the Hamiltonian $H$). The first equation
tells us that the Schwarzschild radius, and thus the mass, is
determined by the entropy deficit. For $R_s \ll R$ we have $\Delta S
= (M/2\pi R)$. Assuming that this relation continues to hold for
less massive localized excitations, which are not black holes, we
derive the Gibbons Hawking thermal spectrum. Finally, we note the
fact that $R_-$ cannot be increased indefinitely: its maximum occurs
when $R_+ = R_- = R/\sqrt{3} $.

It is easy to model these properties in terms of our fermionic
oscillators. Define, in Planck units, $$\pi R^2 \approx {\rm ln 2}
N^2,$$
$$\pi R_{\pm}^2 \approx {\rm ln 2} N_{\pm}^2.$$  More precisely, fix
an integer $N_-$ and define $N_+ $ to be the closest integer
approximation to the solution of
$$N_+ N_- + N_+^2 + N_-^2 = N^2.$$
Define the ensemble of black hole states to be those satisfying

$$\psi_i^A |BH > = 0, \ \ \ i = 1 \ldots N_-  , \ \ \ A = 1\ldots
N_+ .$$ The basis in which this is true is chosen arbitrarily.  By
analogy with our discussion of particle states, we may think of this
as a choice of the horizon volume in which the black hole sits.

The entropy deficit of this ensemble, relative to the maximally
uncertain density matrix is $N_+ N_- {\rm ln} 2 $, which is, for
large $N_-$, what we expect from a Schwarzschild-dS black hole,
given our identification of horizon radii with $N_{\pm}$. We can
then invent a Poincare Hamiltonian $P_0$ such that the black hole
ensemble consists mostly of states whose eigenvalue satisfies the
classical relation between mass and entropy. To do this we note that
the statistical expectation value of the fermion number operator
$${\bf N}  \equiv (\psi^{\dagger})^j_B \psi_j^B,$$ is
$$\langle\langle{\bf N}\rangle\rangle ={1\over 2} (N - N_+
N_-),$$ and its fluctuations are of order $1/N$ for large $N$.

Thus, if
$$P_0 \equiv \sqrt{{{\rm ln} 2}\over 2\pi} M_P (1 -
{{2{\bf N}}\over {N^2}}) \sqrt{N^2 - {\bf N}} ,$$ and we make the
above identifications of integers with radii, then
$$\langle\langle P_0 \rangle\rangle = M .$$
This equation should be understood as what I have elsewhere
\cite{grossbday} called the {\it asymptotic darkness} approximation:
black holes are, in this approximation, degenerate eigenstates of
the high energy limit of the Hamiltonian. The explicit construction
adds a new wrinkle: even in this approximation the states are
non-degenerate, and the black hole energy is a statistical average.
Improvements to the asymptotic darkness approximation will lift the
degeneracy of states of equal ${\bf N}$ and replace it by a random,
closely spaced set of levels with density $2^{ - N_-^2}$. They will
also allow black holes to decay into particle states\footnote{The
phenomenology of Hawking decay means that the width of these states
is much larger than the level density. }, and it is likely that low
entropy members of the approximate black hole ensemble with ${\bf
N}$ far from $\langle\langle {\bf N}\rangle\rangle R$ will be
particles rather than black holes.

The last few sentences were fantasies of hypothetical future work.
What we have accomplished in this section is a construction of
states and a Hamiltonian with the qualitative features of
semi-classical black holes, and we have constructed them out of the
same variables that we used above to construct particle states. It
is reasonably clear that very large black holes, near the Nariai
maximum, will not admit particle excitations, and that groups of
particles with large momenta will naturally merge into black holes.
The details (in which, famously, the devil resides) remain to be
worked out.

\section{Conclusions}

The formalism of holographic space-time is a fully quantum
mechanical system of axioms, in which space-time geometry is an
emergent property of a class of large quantum systems. The causal
structure of the space-time is fixed, but is determined in terms of
possible solutions to an infinite set of dynamical consistency
conditions.  So far, the only known solution of these equations is
the DBHF cosmology of \cite{bfm}. This is a mathematically well
defined model, and forms the starting point for a more realistic
cosmology based on the idea of normal defects in the DBHF. Using the
Israel junction condition and simple scaling arguments, holographic
cosmology provides the first complete theory of the initial
conditions of the universe. In particular, it provides a rationale
for the low initial entropy one must assume in standard cosmology.
The system undergoes a phase transition to a dilute black hole gas
at a certain scale of energy density, well below the Planck scale.
Prior to this transition the formalism of quantum field theory is
not a good approximation to the physics. Just after the transition,
the system is well modeled by a gas of black holes whose size is a
bit smaller than the particle horizon. If the gas is relatively
homogeneous this gas will expand freely and the black holes will
evaporate. If not, it will re-collapse into the DBHF.

The DBHF defect model thus derives the homogeneity, isotropy,
flatness, and low initial entropy of the universe without recourse
to inflation. It cannot however account for the long range
correlations in the CMB data, without a small amount of inflation
(perhaps 20 e-folds), but it does provide the rationale for the
homogeneous initial conditions assumed in most inflationary models.
\footnote{I believe that the standard claim that inflation
automatically explains this (the universe is a free lunch) is based
on a highly un-natural assumption that most of the degrees of
freedom of the current universe, which {\it cannot} be modeled by
quantum field theory in the initial inflationary patch, are in the
ground state of some adiabatic Hamiltonian. This puts in rather than
derives the very special nature of the initial state.}

Finally, the Israel junction condition implies that any region of
space-time that has a normal equation of state in the asymptotic
future, must evolve to de Sitter space. One dS horizon volume is
embedded in the $p=\rho$ DBHF background as a marginally trapped
surface. The c.c. of this asymptotically dS space time is determined
by cosmological initial conditions. It counts the number of degrees
of freedom that have escaped falling into the equilibrated DBHF
fluid. We are led to the concept of a {\it lonely multiverse}, a
universe composed of a distribution of normal cosmologies, all
asymptotically future dS, embedded as marginally trapped surfaces in
a $p=\rho$ background. The {\it a priori} measure on the
cosmological constant favors large values of $\Lambda$, but the
initial amplitude of density fluctuations is bounded\footnote{One of
the quantitative questions that is so far unanswered, is whether
this {\it a priori} bound is close to the observed value of
primordial density fluctuations.}. Anthropic
considerations\cite{btetal} favor smaller values of $\Lambda$, so we
might imagine the observed value is a compromise between these two
criteria\footnote{The hypothetical connection between SUSY breaking
and the c.c., which we discussed, also puts anthropic lower bounds
on $\Lambda$\cite{pentagon}.}.

One may ask whether other parameters of low energy physics have a
similar random distribution\footnote{with the attendant
phenomenological difficulties posed by this
hypothesis\cite{bdmbdg}.}. In the context of holographic space-time,
this is the question of how many different theories of stable dS
space we can construct. It should be noted that whether or not the
string landscape of meta-stable dS spaces exists, none of them can
be the same as the models which one will construct by holographic
methods. The latter have, by construction, a finite number of
quantum states, whereas the former, by virtue of their decays into
zero c.c. regions of moduli space, seem to require an infinite
number.

\section{\bf Appendix}

\subsection{The sound of one hand waving at the horizon}

In the text, I argued for the relation $m_{3/2} = \kappa
\Lambda^{1/4}$ because $N^{1/2} \sim (M_P /\Lambda^{1/4}) $ is the
parameter that controls the convergence of the Hilbert space of
single particle states to its semi-classical limit. $N^{1/2}$ is the
cutoff on single particle angular momenta. In this appendix, I want
to recall another argument for the same scaling.

That argument was based on effective field theory, modified by the
interaction of particles with thermal states on the dS horizon.
These states are not well described by field theory and are almost
exactly degenerate. We recall that the super-Poincare invariant
limiting theory has an R symmetry, which acts on the supercharges
like some $Z_k$ phase with $k \geq 3$. When this symmetry is
unbroken, the low energy effective theory does not have a SUSY
violating state. R symmetry is broken by interactions with the
horizon, and these induce the SUSY violating state which represents
the correct physics in dS space.

Consider the Feynman diagram computation of some R violating term in
the effective Lagrangian near the origin\footnote{Near the horizon,
the static observer sees a very high temperature state in which SUSY
is violently broken.}.  In order to interact with the horizon
degrees of freedom, at least one particle line must propagate out to
the horizon, which is a space-like distance  $\pi R/2$ away. In
order to give an R violating interaction, that particle must carry R
charge and we will assume that the gravitino is the lightest
particle with this property:  thus we have a suppression

$$\delta {\cal L} \sim e^{- \pi m_{3/2} R}$$ from the two gravitino
lines going out to the horizon. The gravitino is absorbed by the
horizon, via some interaction operator $V$ and then re-emitted by
the same operator. Since the horizon states are degenerate we get

$$\delta {\cal L} \sim e^{- \pi m_{3/2} R} \sum_n |\langle g | V | n
\rangle |^2 .$$ Now we ask, which states of the horizon are likely
to give matrix elements of order one?  Like Landau level states, the
degenerate horizon states can be localized on the sphere, with a
number of states proportional to $e^{area}$ of the localized region.

The hand waving part of the calculation consists of the following
statements
\begin{itemize} \item Gravitinos, being massive, can only propagate near
the null horizon for a proper time of order $m_{3/2}^{-1}$.

\item During this proper time, the gravitinos perform a random walk,
with Planck step size, on the horizon, thus covering an area $
{b\over {m_{3/2} M_P}}$.

\end{itemize}

A finite fraction of the states in this area are assumed to have
matrix elements of order one. Thus

$$\delta {\cal L} \sim e^{ - \pi m_{3/2} R} e^{{{b M_P}\over
{m_{3/2}}}} .$$

If $m_{3/2}$ is to have any power law behavior at all when $R M_P
\rightarrow\infty$, then the positive and negative exponentials must
cancel exactly:

$$ \pi m_{3/2} R = {{b \over {m_{3/2} M_P}}}.$$ Plugging in the
relation of $R$ and the cosmological constant we get
$$m_{3/2} = ({{8 b^2} \over {9 \pi}})^{1/4}  \Lambda^{1/4}.$$
This gives a formula for the unknown constant $\kappa$ in terms of
the unknown constant $b$.

\section{Acknowledgments}

I would like to thank Sasha Zamolodchikov, Sasha Belavin and
Y.Pugai, for inviting me to the conference where this material was
presented.  I also wish that I could thank Alyosha Zamolodchikov for
all of the inspirational physics he left to the world, and for his
friendship over the years. He will be sorely missed.

This research was supported in part by DOE grant number
DE-FG03-92ER40689.

\end{document}